# How to Mitigate the Dependencies of ChatGPT-4o in Engineering Education

Maoyang Xiang, T. Hui Teo*
Engineering Product Development
Singapore University of Technology and Design
*Corresponding Author: tthui@sutd.edu.sg
These authors contributed equally to this work.

## Abstract

The rapid evolution of large multimodal models (LMMs) has significantly impacted modern teaching and learning, especially in computer engineering. While LMMs offer extensive opportunities for enhancing learning, they also risk undermining traditional teaching methods and fostering excessive reliance on automated solutions. To counter this, we have developed strategies within curriculum to reduce the dependencies on LMMs that represented by ChatGPT-4o. These include designing course topics that encourage hands-on problem-solving. The proposed strategies were demonstrated through an actual course implementation. Preliminary results show that the methods effectively enhance student engagement and understanding, balancing the benefits of technology with the preservation of traditional learning principles.

Keywords: computer engineering, hardware description language, large multimodal models.

## Introduction

In recent years, the expansive application of LMMs across numerous sectors has been particularly notable in education, where they serve multiple roles. LMMs can enhance personalized learning by adapting content to individual student needs, providing immediate feedback on assignments, and supporting diverse learning styles. They also facilitate the creation of interactive and dynamic educational materials, making complex subjects more accessible through simulations and explanatory models. In the instructor support, these models can automate routine tasks such as grading and responding to common student inquiries, allowing educators more time to focus on in-depth teaching and student interaction.

The widespread application of LMMs in education has introduced challenges to the traditional teaching system. One significant issue is the disruption of the traditional learning link, where students are increasingly turning to LMMs for assistance with homework and exercises rather than completing these tasks independently. While this approach broadens their knowledge base, it often undermines their development of critical thinking skills. More critically, it distorts the long-established learning paradigm of input-output evaluation. This issue is particularly pronounced in engineering disciplines, which require extensive hands-on practice to build experience and expertise.

An actual course delivery experience is shared in the paper. The course name is Digital System Laboratory, which is a senior year undergraduate course that covers digital circuits and system hardware design. In this Digital System Laboratory (DSL) course, a concerning trend has emerged: many students submit assignments that rely heavily on content generated by LMMs, and the proportion of such submissions is increasing. Concurrently, these students demonstrate a noticeable lack of actual engineering skills during practical tasks. The root cause is the overuse of LMMs, which masks the students' gaps in knowledge and engineering abilities, leading to superficial understanding rather than deep learning.

To counteract the negative impact of LMMs, our DSL course has implemented several innovative strategies. We have redesigned assignments and tasks to be more aligned with real-world engineering practice rather than traditional textbook exercises. Students are required to analyze problems in the context of actual engineering needs, fostering the development of critical thinking skills and ensuring a more comprehensive understanding of the subject matter.

This paper begins by outlining the challenges posed by the integration of LMMs in educational settings, particularly in engineering disciplines. Following this, the paper presents the strategies implemented to mitigate these effects, emphasizing the importance of hands-on, practice-based learning. The conclusion synthesizes these findings, offering insights and recommendations for educators seeking to balance the benefits of LMMs with the need for traditional, foundational learning.

# Methods

Games Theory is adopted as the baseline of the methodology of using LMMs in the classroom in this work. It is interesting to experience how Colonel Blotto game theory [1] that dealing with attack-defense scenarios, find a place in classroom due to LMMs. At the attack part, the learners trained their LMMs based on the course materials including homework questions/solutions, [2, 3] and use the trained LMMs to help them in answering questions including open-book questions during quizzes. At the defense part, the instructors also use LMMs to help in improving the assignment, quizzes questions to be LMMs proof.

This section provides an overview of the DSL course and details the methods employed when setting up homework and quizzes.

## Digital System Laboratory Overview

DSL course offers students a hands-on learning experience in the fundamentals of digital systems and their applications. This course covers the design, analysis, and implementation of digital circuits and systems, including logic gates and flip-flops. Through these activities, the course aims to develop critical thinking and problem-solving skills, preparing students for advanced studies or careers in electronics and computer engineering.

In the DSL course, students are expected to implement digital systems using Hardware Description Languages (HDL), [2, 3]. Their journey begins with understanding basic digital circuits and mastering the corresponding HDL syntax to develop good coding habits. Initially, students engage in exercises where they use HDL to simulate basic logic gates. Subsequently, these gates are combined to create more complex digital systems. However, the rise of LMMs has disrupted this traditional learning path. LMMs can effortlessly solve basic training problems, but often at the cost of poor coding style and habits. This leads to a scenario where beginners are ill-prepared for designing complex digital systems in the future, as they miss out on the foundational practices essential for their development. Therefore, in setting the course assignments, efforts should be made to avoid problems for which LMMs can directly generate corresponding answers.

# Real Application Hands On

The current challenge in designing course assignments is to strike a balance between ensuring that LMMs cannot directly provide answers and guaranteeing that students receive sufficient basic practice. In the DSL course, we address this dilemma by incorporating practical engineering concepts into fundamental topics. This approach creates complexities that LMMs struggle to interpret, while students with basic knowledge can still comprehend the problem requirements.

A basic HDL question, a counter with reset function with and without the approach, is proposed for demonstration. The following is examples of traditional HDL question from textbook:

> Design a 4-bit synchronous up counter in Verilog HDL that counts from 0 to 9. Your design should include the following components:
> **Counter Logic**: Implement the counter logic to increment the count value on each positive edge of the clock. If the reset is low, the counter value should reset to 0, irrespective of the clock's state.
> **Counter NUM Output**: Invert the counter value and connect it to the NUM port.
> **Count Limit and Output Signal**: Ensure that the counter value resets to 0 when it reaches 10, and the cout signal should be high only when the counter is at 9.
>
> Code Template:
> ```
> module counter10(
>     input reset,
>     input clock,
>     output [3:0] NUM,
>     output COUT
> );
> endmodule
> ```

The basic application circuit is combined with the counter module to generate a new question as following:

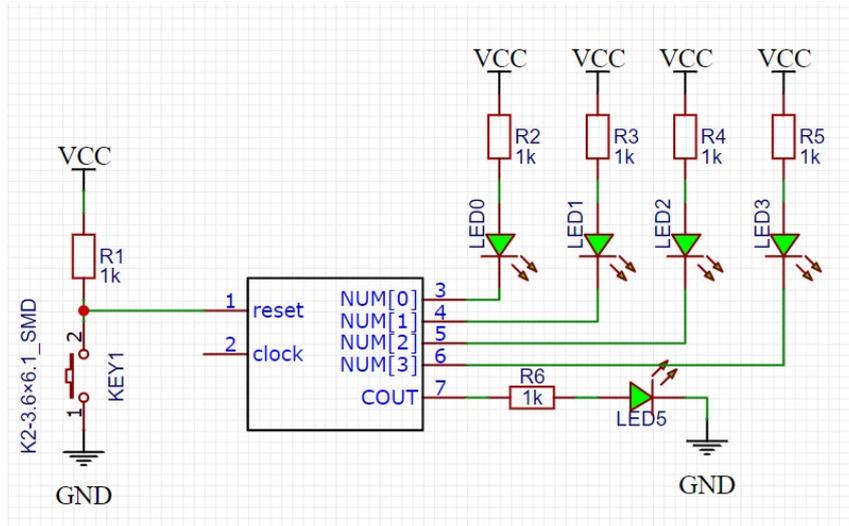

Verilog Question: A 4-bit counter, named counter10, is designed to perform three key functions: reset, count, and overflow indication. The specific circuit block diagram is illustrated below:

Functionality Overview:

1. Reset Function:

**Trigger**: When KEY1 is pressed.

**Effect**: Resets counter10, ensuring that LEDs from LED3 to LED0 are turned off, and the overflow indicator LED5 is also deactivated.

2. Counting Mechanism:

**Trigger**: Upon the release of KEY1, the counter reacts to the falling edge of the clock signal.

**Initial Output**: At the first falling edge of clock, LED0 is activated, while LEDs LED3 to LED1 remain off.

**Sequential Counting**:

- For each subsequent falling edge of the clock, the counter increments by one, which should be displayed by LED3, LED2, LED1, LED0.
- This pattern proceeds until the counter displays 1001. At this state, LED3 and LED0 are on, while LED1 and LED2 are off. Concurrently, the overflow indicator LED5 is activated.

**Final State**: Upon the next falling edge of the clock after reaching 1001, all LEDs, including the overflow indicator LED5, are turned off.

Code Template:

```
module counter10(
    input reset,
    input clock,
    output [3:0] NUM,
    output COUT
);
endmodule
```

Both questions aim to create the same counter. However, the second question places the counter within a real application circuit, complete with a schematic, and describes the module's behaviour in an indirect manner. Students with basic electronics knowledge can easily translate the circuit into counter requirements, which presents a moderate challenge for modern LMMs due to the use of schematics and the indirect description of behaviour.

By integrating real-world engineering constraints and context into this basic exercise, we create a scenario where students must apply their foundational understanding in a practical setting, effectively navigating around the limitations of LMMs in providing direct solutions.

# Results

Table 1. Result of different LLM with two questions.

| LMMs | Textbook Question | Purposed Question |
|---|---|---|
| GPT-3.5 | PASS | Synthesizable |
| GPT-4 | PASS | Synthesizable |
| GPT-4o | PASS | Synthesizable |
| Copilot (Microsoft) | PASS | Synthesizable |

The effectiveness of the questions was evaluated using four common LLM models, and the outcomes are presented in Table 1. The analysis was conducted on May 21, 2024, and it's important to note that results may vary based on the test date. The findings demonstrate that modern LMMs can successfully handle straightforward textbook HDL questions. However, for the proposed question, while the models produced synthesizable results, they did not fully meet the question's requirements.

The results from the testing of four common LMMs on two types of HDL questions reveal that while modern LMMs are adept at solving straightforward textbook questions, they struggle with more complex, application-oriented questions. Specifically, the LMMs could generate synthesizable code for the proposed question but failed to fully capture the nuances and specific demands of the question. This discrepancy underscores the need for question designs that challenge LMMs and push for deeper understanding and application in students, rather than relying on models that excel primarily at textbook-style problems.

# Discussion

Initial results from implementing these strategies in the DSL curriculum have been promising. Students show increased engagement and a deeper understanding of course material. Feedback indicates that balancing technology with traditional learning principles enhances the overall educational experience.

While LMMs hold great promise for transforming education, their integration into the curriculum must be managed carefully to avoid undermining traditional teaching methods. The strategies we have developed for the DSL curriculum demonstrate that it is possible to leverage the benefits of LMMs while preserving the fundamental principles of hands-on, experiential learning. However, the development and use of LMMs in education remain challenging, requiring ongoing adaptation and refinement to ensure they enhance rather than detract from the learning process.

Future research and development should focus on refining these strategies, exploring new methods to integrate LMMs effectively into various educational contexts, and continuously assessing their impact on student learning outcomes. By doing so, we can ensure that LMMs serve as valuable tools that complement and enhance traditional educational practices.

# Acknowledgments

We would like to thank SUTD-ZJU IDEA Visiting Professor Grant (SUTD-ZJU (VP) 202103, and SUTD-ZJU Thematic Research Grant (SUTD-ZJU (TR) 202204), for supporting this work.